\documentclass[sigconf]{acmart}

\AtBeginDocument{%
  }

\setcopyright{acmlicensed}

\copyrightyear{2025} 
\acmYear{2025} 
\setcopyright{acmlicensed}\acmConference[CHI '25]{CHI Conference on Human Factors in Computing Systems}{April 26-May 1, 2025}{Yokohama, Japan}
\acmBooktitle{CHI Conference on Human Factors in Computing Systems (CHI '25), April 26-May 1, 2025, Yokohama, Japan}
\acmDOI{10.1145/3706598.3713852}
\acmISBN{979-8-4007-1394-1/25/04}

\usepackage{amsmath}
\usepackage{multirow}
\usepackage{relsize}
\usepackage{soul}
\usepackage{subcaption}
\usepackage{xcolor}

\begin{document}

\title{Exploring Personalized Health Support through Data-Driven, Theory-Guided LLMs: A Case Study in Sleep Health}

\author{Xingbo Wang}
\orcid{0000-0001-5693-1128}
\affiliation{%
  \institution{Weill Cornell Medicine, \\Cornell University}
  \city{New York}
  \state{NY}
  \country{USA}
}
\email{xiw4011@med.cornell.edu}
\email{wangxbzb@gmail.com}

\author{Janessa Griffith}
\orcid{0000-0003-2629-2905}
\affiliation{%
  \institution{Cornell Tech}
  \city{New York}
  \state{NY}
  \country{USA}
}
\email{janessa.griffith@cornell.edu}

\author{Daniel A. Adler}
\orcid{0000-0003-3328-0312}
\affiliation{%
  \institution{Cornell Tech}
  \city{New York}
  \state{NY}
  \country{USA}
}
\email{daa243@cornell.edu}

\author{Joey Castillo}
\orcid{0009-0009-5459-0345}
\affiliation{%
  \institution{Cornell Tech}
  \city{New York}
  \state{NY}
  \country{USA}
}
\email{jc3292@cornell.edu}

\author{Tanzeem Choudhury}
\orcid{0000-0002-5952-4955}
\affiliation{%
  \institution{Cornell Tech}
  \city{New York}
  \state{NY}
  \country{USA}
}
\email{tanzeem.choudhury@cornell.edu}

\author{Fei Wang}
\orcid{0000-0001-9459-9461}
\affiliation{%
  \institution{Weill Cornell Medicine}
  \city{New York}
  \state{NY}
  \country{USA}
}
\email{few2001@med.cornell.edu}

\renewcommand{\shortauthors}{Wang et al.}

\newcommand{\argmax}{\operatornamewithlimits{arg\,max}}

\newcommand{\xingbo}[1]{{\color{black} #1}}
\newcommand{\janessa}[1]{{\color{orange} #1}}
\newcommand{\dan}[1]{{\color{cyan} #1}}
\newcommand{\todo}[1]{{\color{red} #1}}
\newcommand{\jianben}[1]{{\color{black} #1}}
\newcommand{\rev}[1]{{\color{black} #1}}
\newcommand{\revv}[1]{{\color{black} #1}}

\newcommand{\ie}{i.e.}
\newcommand{\eg}{e.g.}
\newcommand{\esp}{esp.}
\newcommand{\etal}{et al.}
\newcommand{\imp}[1]{\textbf{\textit{{#1}}}}
\newcommand{\systemname}{\textsf{{\MakeUppercase{HealthGuru}}}}
\newcommand{\name}{\textsf{\MakeUppercase{H}\smaller\MakeUppercase{ealth}\larger\MakeUppercase{G}\smaller\MakeUppercase{uru}}}


\begin{abstract}
Despite the prevalence of sleep-tracking devices, many individuals struggle to translate data into actionable improvements in sleep health. Current methods often provide data-driven suggestions but may not be feasible and adaptive to real-life constraints and individual contexts. We present \name{}, a novel large language model-powered chatbot to enhance sleep health through data-driven, theory-guided, and adaptive recommendations with conversational behavior change support. \name{}'s multi-agent framework integrates wearable device data, contextual information, and a contextual multi-armed bandit model to suggest tailored sleep-enhancing activities. The system facilitates natural conversations while incorporating data-driven insights and theoretical behavior change techniques. Our eight-week in-the-wild deployment study with 16 participants compared \name{} to a baseline chatbot. Results show improved metrics like sleep duration and activity scores, higher quality responses, and increased user motivation for behavior change with \name{}. We also identify challenges and design considerations for personalization and user engagement in health chatbots.
\end{abstract}

\ccsdesc[500]{Human-centered computing~Interactive systems and tools}
\ccsdesc[500]{Human-centered computing~HCI design and evaluation methods}
\ccsdesc[500]{Human-centered computing~Natural language interfaces}
\ccsdesc[300]{Computing methodologies~Natural language processing}
\keywords{Personalized Health, Sleep Health, Behavior Change, Large Language Models, Activity Recommendations}


\maketitle

\section{Introduction}\label{sec.intro}

Sleep is a fundamental pillar of human health, profoundly influencing physical well-being, cognitive function, and emotional resilience~\cite{medic2017short}. 
With modern wearable devices like smartwatches and rings, monitoring sleep patterns has become more accessible than ever. These technologies provide detailed metrics—sleep duration, stages, heart rate variability, and more—offering valuable insights into nightly rest. However, they often fall short in delivering actionable guidance.
Users are frequently presented with an abundance of metrics but lack clear guidance on how to interpret and apply this data to improve their sleep quality.

Many systems have aimed to bridge this gap through personalized sleep interventions, but most have shown significant limitations. 
Commercial devices like Fitbit, Whoop, and the Oura Ring, as well as research prototypes like SleepCoach~\cite{daskalova2016sleepcoacher}, provide detailed insights into sleep patterns by correlating sleep factors with sleep quality but often offer generic recommendations that rely on predefined static templates. These templates are not flexible enough to adapt to users' evolving preferences, schedules, or environmental factors. 
SleepGuru~\cite{lee2022sleepguru} incorporates users' calendars to predict sleep pressure and recommends sleep schedules accordingly. 
However, it falls short in addressing crucial real-life variables that extend beyond calendar events, such as physical activity or changes in personal habits.

Conversational chatbots have emerged as a promising interaction paradigm for eliciting user personal contexts and preferences and delivering personalized health interventions~\cite{bickmore2005establishing, fitzpatrick2017delivering}.
However, many health-focused chatbots reply on template-based or rule-based methods, limiting their conversational flexibility and their ability to provide dynamic personalization~\cite{laranjo2018conversational}. 
This rigidity can lead to repetitive or irrelevant suggestions, potentially reducing engagement and motivation for behavior change over time.

Recent advances in large language models (LLMs) have shown great potential for enabling more flexible, context-aware conversations.
However, most existing LLM applications in health are still in their early stages. 
They struggle to natively support processing and interpreting raw and complex time-series sensor data, as well as translating data insights into actionable plans tailored to users' personal contexts. This gap hinders their effectiveness in driving sustained behavior change in a structured and systematic manner.

To address these identified gaps, we present \name{}, a novel LLM-powered chatbot augmented by a multi-agent framework to provide personalized, data-driven, and theory-guided sleep health support. 
\name{} seamlessly integrates wearable device data, contextual information, and established behavior change theories to deliver adaptive recommendations and motivational support. 
To tackle the challenge of delivering context-aware, adaptive interventions, 
we first build a contextual multi-armed bandit (MAB) model that dynamically learns and suggests sleep-enhancing activities by considering real-time environment factors (\eg, time and weather) and individual physiological sleep data collected from wearable devices.
Informed by behavioral theories~\cite{bandura1977self,lally2010habits}, this model balances exploiting previously successful sleep-enhancing activities and exploring new, potentially beneficial options. This ensures that recommendations remain relevant and adapt to users' evolving contexts and preferences.
To enhance real-time adaptability and intervention effectiveness, we leverage LLMs that incorporate behavior change techniques to deliver the activity recommendations in a natural and motivational conversation.
Combining the strengths of LLMs and the contextual MAB model while overcoming their limitations, we design a multi-agent framework where specialized agents coordinate LLMs' conversational and reasoning capabilities with context-aware dynamic recommendations. 
These agents work collaboratively to interpret and integrate wearable data insights with context information and deliver adaptive recommendations tailored to users' unique circumstances in a theory-guided conversation.
To evaluate \name{}, we conduct an eight-week in-the-wild deployment study with 16 participants, comparing our system to a baseline chatbot without context-aware recommendations or theory-guided conversations.
Our study adopts a within-subjects design.
The results show significant improvements in sleep metrics (\eg, longer sleep duration) and better activity scores, when using \name{} compared to the baseline. Additionally, participants report higher levels of motivation to engage in sleep-enhancing behaviors, attributed to the personalized and context-aware recommendations provided by the system.

The major contributions of this work are:
\begin{itemize}
    \item We introduce a novel LLM-powered chatbot with a multi-agent framework that integrates wearable data analysis, contextual factors, and a multi-armed bandit model to deliver personalized, timely, and adaptive sleep recommendations guided by behavior change theories.
    \item We conduct a eight-week in-the-wild deployment study to evaluate the effectiveness and usability of \name{} compared to a baseline, showing improvements through sleep metrics and user feedback.
    \item We provide design implications for future personalized health system designs and LLM applications in health applications. 
\end{itemize}

\section{Background and Related Work}\label{sec.rw}

\subsection{Sleep Sensing and Tracking}\label{subsec.sleep_sensing}

Lullaby~\cite{kay2012lullaby} uses multiple sensors (\eg, sound, temperature) to assess and visualize environmental and sleep data together, helping users understand their sleep. 
WAKE~\cite{pham2020wake} is a behind-the-ear device that detects microsleep. FitBit\footnote{\url{https://www.fitbit.com/global/us/technology/fitbit-app}} tracks sleep duration and stages (light, deep, REM) and provides overall sleep quality assessments. Whoop\footnote{\url{https://www.whoop.com/us/en/thelocker/introducing-whoop-coach-powered-by-openai/}} monitors fitness, sleep, and recovery metrics, offering recommendations to improve human performance. 
The Oura Ring\footnote{\url{https://ouraring.com/oura-experience}} collects precise body signals (\eg, temperature, heart rate variability, movement) to generate readiness, sleep, and activity scores, along with personalized suggestions. 
We chose the Oura Ring for our research due to its high-quality sleep measurements and activity detection, which align well with our goals for personalized sleep-enhancing recommendations.

\subsection{Sleep Feedback and Recommendation}\label{subsec.sleep_feedback}
Sleep feedback and recommendation systems generally focus on two key aspects:
1) improving sleep hygiene awareness and 2) providing personalized recommendations based on user lifestyles to improve sleep health.

To improve sleep awareness, systems use various visualization approaches.
ShutEye~\cite{bauer2012shuteye} displays activity timelines on mobile wallpapers to indicate optimal timing for sleep-impacting activities. 
SleepExplorer~\cite{liang2016sleepexplorer} is a web-based tool that uses data visualizations to show the impacts of different contextual factors on sleep quality, such as steps, calories, and water consumed.
SleepTight~\cite{choe2015sleeptight} is a mobile application that further supports self-reporting on influential sleep factors and their impacts on sleep.

To provide personalized sleep recommendations, many studies incorporate user modeling techniques that consider user characteristics, behaviors, and contexts.
SleepCoach~\cite{daskalova2016sleepcoacher} utilizes static user models to correlate sleep factors with predefined recommendation templates of general guidelines, such as ``For the next 6 days, try going to bed at 11 PM.'' 
\citet{daskalova2018investigating} improved personalization by clustering users into groups based on shared physical profiles or recommendations.
However, this cohort-based model is inherently static and may overlook the nuances of individual behaviors and preferences over time.
\citet{lee2022sleepguru} built a dynamic model that considers users’ daily physiological data and calendar events to predict sleep pressure and sleepiness, which are used to recommend sleep schedules. However, it only focuses on calendar events, which constrains their applicability to broader lifestyle contexts and users' preferences.
Building upon these studies, we adopt a dynamic user modeling approach that incorporates just-in-time adaptive intervention (JITAI) principles~\cite{nahum2018just}.
JITAI is a design methodology that aims to deliver timely and personalized support by dynamically adjusting to users' internal and contextual conditions when individuals are in need of health support and receptive to engaging with an intervention.
To operationalize JITAI for sleep health, we build a contextual multi-armed bandit model inspired by previous reinforcement learning approaches~\cite{mybehavior,liao2020personalized} that personalizes interventions by learning from user behaviors over time.
The model, informed by self-efficacy~\cite{bandura1977self} and habit formation~\cite{lally2010habits} theories, dynamically adjusts recommendations by balancing the exploitation of proven successful sleep-enhancing activities with the exploration of novel ones.
Our model considers both immediate contextual factors (\eg, time and weather) and long-term user physiological data from wearable devices.
Furthermore, to enhance real-time adaptability and intervention effectiveness, we deliver these recommendations through a theory-guided LLM-based chatbot, which integrates behavior change techniques to address users' needs and preferences.

\subsection{LLMs for Health Applications}\label{subsec.llms4health}
\citet{kim2022leveraging} leveraged LLMs to streamline natural language interactions that can understand users' utterances and automatically generate questions for users to co-build retrospective activity logs for self-tracking.
\citet{englhardt2023classification} systematically investigated the use and evaluation of LLMs for multi-sensor data classification and reasoning to facilitate the clinical workflows of therapists.
\citet{abbasian2023conversational} presented a personalized LLM-powered framework for conversational health agents.
The framework aims to deliver personalized responses to users' healthcare-related queries via the analyses of users' questions and relevant and essential data from different external sources.
Recently, PH-LLM~\cite{cosentino2024towards}, a version of Gemini, has been fine-tuned for personal health applications in sleep and fitness, demonstrating performance comparable to human experts on long-form case studies and professional exams.
PHIA~\cite{merrill2024transforming} is an LLM-based agent framework that uses iterative reasoning, code generation, and web search to analyze and interpret personal health data from wearables.
However, all these LLMs and systems focus on static screenshots of user data.
Building upon prior work, we build an LLM-powered chatbot. 
This chatbot integrates an activity recommendation model that analyzes dynamic users' sleep and activity patterns and offers personalized advice or answers specific questions about their sleep health.
Moreover, by employing a behavior change framework, the chatbot motivates users to adopt recommended activities, thereby fostering better sleep habits.

\section{HealthGuru}\label{sec.bg}
\name{} is a novel personalized health support chatbot designed to promote behavior change for better sleep health. By integrating wearable data, contextual information, and established behavior change theories, it offers data-driven, theory-guided support through an LLM-based multi-agent conversational interface.

\subsection{Data-Driven and Theory-Guided Health Support}

In designing \name{}, we focused on four key requirements:

\begin{itemize}
     \item \textbf{Provide quantitative health data analytics.} 
     Wearable devices facilitate health monitoring by providing rich, longitudinal data streams~\cite{piwek2016rise}. \name{} integrates relevant wearable data to provide a holistic view of users' health status, offering tangible metrics to track sleep and activities while providing nuanced, data-driven health advice.
 
    \item \textbf{Adapt to personal contexts and preferences.} 
    Health behaviors are deeply influenced by environmental and personal contexts~\cite{sallis2015ecological}.
    Static recommendations often fail to capture the dynamic nature of users' lives~\cite{yardley2016understanding}.
    Guided by just-in-time adaptive intervention (JITAI) principles, our system delivers personalized and timely support by considering users' realtime internal states (\eg, physiological metrics) and external contexts (\eg, environment)~\cite{nahum2018just}.
    We employ a contextual multi-armed bandit approach to dynamically adapt recommendations to users' evolving health states (\eg, physiological metrics) and external contexts (\eg, time, weather).
    This ensures the relevance and effectiveness of health interventions in changing environments. 

    \item \textbf{Support theory-informed motivational practice.} 
    While providing health information and knowledge is essential, it is often insufficient on its own to drive sustained behavior change~\cite{kelly2016changing}.
    Established behavioral change theories~\cite{michie2013behavior, atkins2017guide} offer structured approaches to understanding and influencing health behaviors.
    Therefore, we ground our \jianben{system-generated} advice on the theoretical techniques and frameworks (\eg, positive reinforcement and goal-setting) to motivate users in their health improvement journey.

    \item \textbf{Enable natural conversational interaction.} 
    The mode of interaction significantly impacts user engagement and the effectiveness of digital health interventions~\cite{bickmore2005establishing}. 
    Conversational interactions have been shown to provide human-like interaction, and increase user satisfaction, trust, and adherence in health applications~\cite{laranjo2018conversational, kocaballi2019personalization}. Through an LLM-based chatbot, \name{} facilitates nature and engaging interactions that can deliver JITAI to address users' immediate sleep health concerns and improve user receptivity to interventions according to user feedback.
   
\end{itemize}

\subsection{System and Data}\label{sec.system_data}
\name{} is built on a multi-agent architecture (\autoref{fig:system_framework}) that processes user inputs, integrates contextual data, and generates personalized, theory-guided responses to promote behavior change and improve sleep health. 
This section outlines the system workflow and the data types used for personalized feedback.

\subsubsection{System Workflow}
\begin{figure*}
  \includegraphics[width=\textwidth]{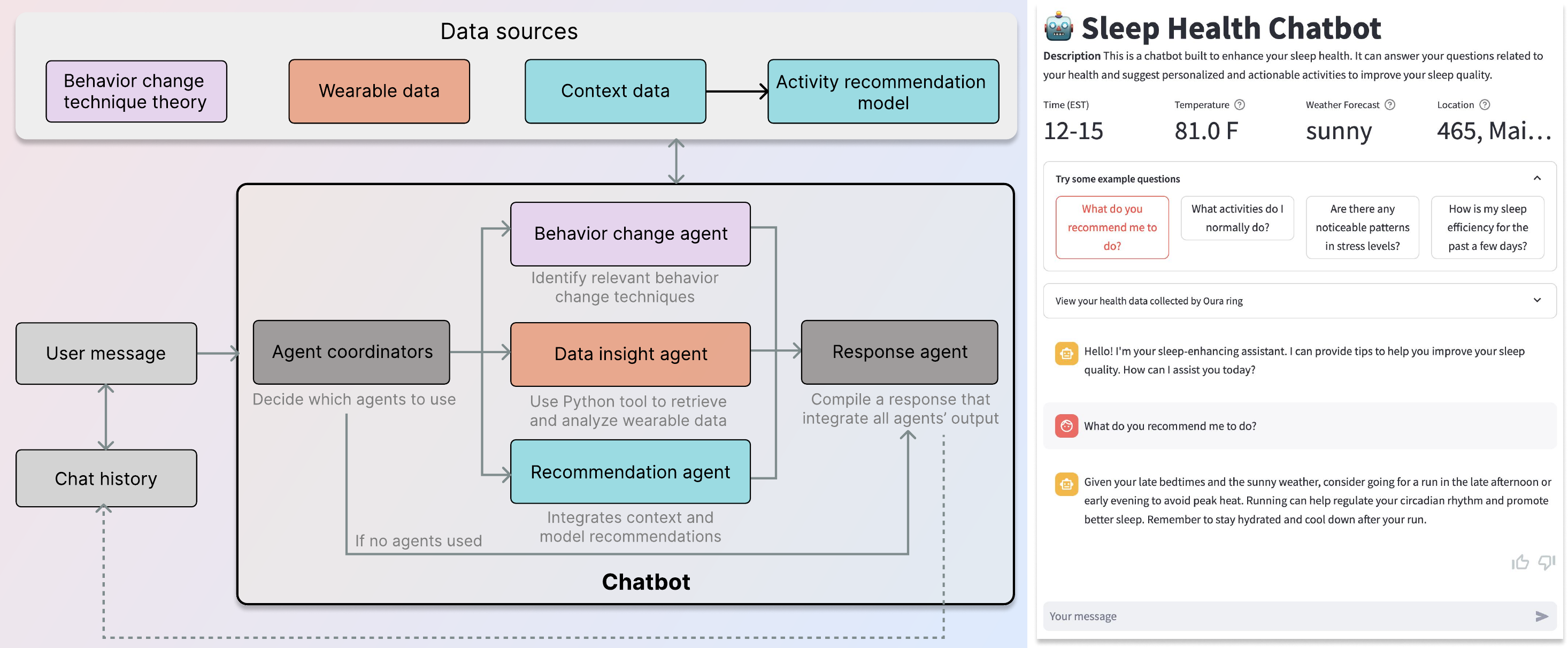}
  \caption{Overview of \name{} system. The system framework (left) features a multi-agent architecture, incorporating behavior change techniques, wearable data analysis, and context-aware recommendations. 
  The user interface (right) engages users in multi-turn conversations to gain insights into their sleep and activity data and receive personalized advice.}
  \label{fig:system_framework}
\end{figure*}

The system processes user messages and chat history by engaging specialized agents that consider behavior change theory, wearable data, context data, and activity recommendations.
When a user sends a message (\eg, ``What do you recommend me to do?''), the agent coordinator selects the appropriate agents. For example, the behavior change agent identifies motivational techniques, while the recommendation agent uses current context (\eg, sunny weather at 81°F) to suggest suitable activities. 
The response agent then compiles these inputs into a concise, coherent reply (as shown in \autoref{fig:system_framework}, where a late afternoon run is suggested to avoid peak heat, with benefits explained).

\subsubsection{Data Collection \& Preparation}
As summarized in \autoref{tab:data_collection}, we collect three primary types of wearable data for sleep health analysis and activity recommendation: physiological, sleep, and activity data.
The selection of these data types is based on their direct relevance to both sleep quality and physical activity and their widespread availability in existing commercial wearable devices (\eg, Oura ring, Fitbit, and Whoop).
Moreover, the data selection is supported by extensive research in sleep science~\cite{shaffer2017overview, ohayon2017national, kredlow2015effects, peake2018critical} and human-computer interaction~\cite{liang2016sleep}.
Physiological data (\eg, HRV)~\cite{shaffer2017overview}
offer quantitative insights into sleep quality and activity intensity.
Sleep data details sleep patterns, efficiency, and quality~\cite{ohayon2017national}.
Activity data captures the type, intensity, and duration of physical activities~\cite{kredlow2015effects}.
We also include context data for generating activity recommendations (in \autoref{subsec.act_rec}) and behavior change theory components (in \autoref{sec.chatbot}) for formulating responses.

\begin{table}[t]
\caption{Collected wearable data types and definitions.}
\label{tab:data_collection}
\small
\begin{tabular}{@{}p{0.3\columnwidth}p{0.65\columnwidth}@{}}
\toprule
\textbf{Category --- Measurement} & \textbf{Definition} \\
\midrule
\multicolumn{2}{@{}l@{}}{\textit{Physiological Data}} \\
Average Heart Rate  & Mean heart rate (in bpm) measured during sleep or activity to assess cardiovascular health. \\
Lowest Heart Rate   & The lowest heart rate (in bpm) recorded during sleep—often occurring in deep sleep stages. \\
Average HRV         & Mean heart rate variability (in ms) during sleep; higher values (e.g., 50–100 ms) suggest better stress resilience. \\
Stress Level        & A computed metric based on physiological markers (such as HRV), typically scaled (e.g., 1–100). \\
\midrule
\multicolumn{2}{@{}l@{}}{\textit{Sleep Data}} \\
Bedtime Start \& End & Timestamps indicating when sleep begins and ends (e.g., 2024-07-26T00:26:28-04:00). \\
Day                 & The date corresponding to the sleep data (e.g., 2024-07-26). \\
Sleep Efficiency    & Ratio of sleep time to time in bed, expressed as a percentage (1–100). \\
Readiness Score     & A daily readiness score (1–100) derived from sleep quality and physiological measures. \\
Time in Bed         & Total time spent in bed (in seconds), including periods of wakefulness. \\
Total Sleep Duration& Actual sleep time excluding wake periods (in seconds). \\
Sleep Score         & Overall sleep quality score (1–100) based on duration, efficiency, and physiological data. \\
\midrule
\multicolumn{2}{@{}l@{}}{\textit{Activity Data}} \\
Activity Type       & Type of physical activity (e.g., Running, Walking, Cycling). \\
Intensity           & Activity intensity level (e.g., easy, moderate, hard). \\
Day                 & The date on which the activity was recorded. \\
Start \& End Time   & Timestamps marking the beginning and end of the activity period. \\
\bottomrule
\end{tabular}
\end{table}

\subsection{Physical Activity Recommendations}
\label{subsec.act_rec}
\begin{figure*}
  \includegraphics[width=\textwidth]{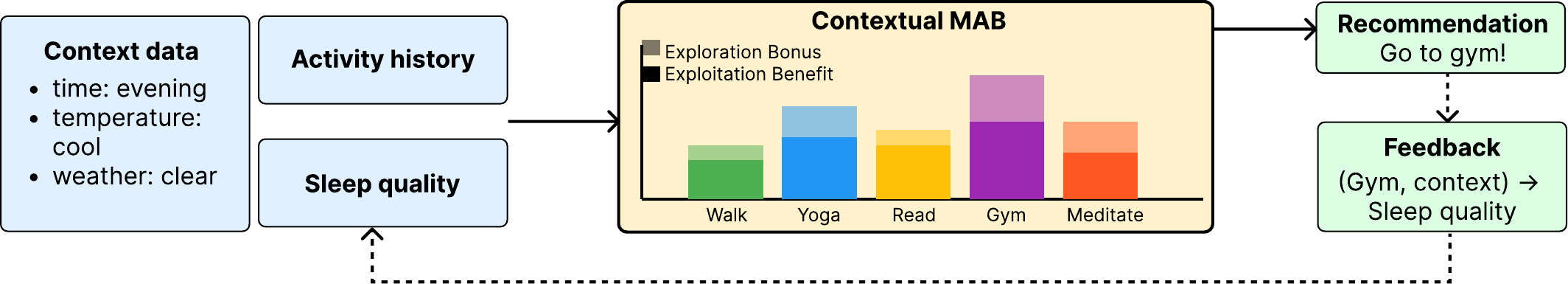}
  \caption{Contextual MAB algorithm workflow. The algorithm generates activity recommendations by balancing the exploration and exploitation benefits of activities based on the current context and the corresponding activity history and sleep scores. After the recommendation, the sleep quality for the recommended activities will provide feedback to update the model.}
  \label{fig:context_mab}
\end{figure*}

We present a novel contextual multi-arm bandit (MAB) model (\autoref{fig:context_mab}) to recommend personalized activities to improve health outcomes.
This approach emphasizes adaptability to user behavior and environmental context factors while maintaining a balance between familiar and novel experiences to motivate behavior change.

\textbf{Context consideration.}
We consider the time of day, temperature, and weather \cite{hardeman2019systematic, klenk2012walking, herbolsheimer2016physical, giles2002relative}.
These factors are crucial in determining the activity feasibility and effectiveness. 
For example, the time of day affects user schedules and energy levels. 
Temperature impacts the comfort of outdoor activities.

\textbf{Problem formulation.}
Framed as an optimization task, the system aims to maximize sleep quality scores (measured by devices like the Oura ring) by recommending activities (\eg, Gym, Walking, Yoga, Reading, Meditation) based on current contexts.

\textbf{Motiviation of contextual multi-arm bandit model}
for activity recommendation is guided by both theoretical and practical considerations.
Self-efficacy theory \cite{bandura1977self} suggests users are more likely to engage in activities they believe they can successfully perform based on their past experience, while habit formation theory \cite{lally2010habits} emphasizes the importance of consistent positive outcomes.
Practically, the system must address the ``cold start'' problem of limited initial user data and adapt to changing user preferences and environmental context over time.
It also needs to balance between recommending known effective and introducing novel options to prevent user boredom and discover new beneficial behaviors.

To address these requirements, we adopt a contextual multi-arm bandit (MAB) model to guide our activity recommendations. In this framework, each ``arm'' represents a physical activity and the ``reward'' is the corresponding sleep quality, while additional context (\eg, time, location, weather) informs decision-making. 
The algorithm balances exploitation (selecting proven activities) with exploration (trying new or less certain ones), aligning with behavioral theories by recommending feasible, context-appropriate options and fostering habit formation through novel experiences. 
The model's adaptive learning capability allows it to respond to changing user preferences and environmental conditions, while also addressing the cold start problem by quickly learning from limited initial data.

\textbf{Implementation of contextual MAB}.
We implement the Linear Upper Confidence Bound (LinUCB) algorithm~\cite{li2010contextual}
for its ability to handle context information and enable online learning with strong theoretical performance.
The core idea behind LinUCB is to model the expected reward of each action as a linear function of the contextual features and incorporate an uncertainty term to balance exploration and exploitation. 
Considering both estimated rewards and confidence in the estimates, the model selects the actions that have the highest potential for reward.
The implementation of LinUCB is described as follows:
\begin{itemize}
    \item \textbf{Context vector construction:} For each recommendation at time $t$, we create a context vector $x_t \in \mathbb{R}^d$, where $d$ is the dimensionality of the context features. The context vector incorporates relevant factors, including time of day, temperature, and weather conditions. Time of day is discretized into seven intervals based on hours: `0-6', `6-9', `9-12', `12-15', `15-18', `18-21', and `21-24'.
    Temperature values are converted into `cold', `mild', `warm', and `hot'.
    Weather is categorized into `sunny',
    `rain', `clear', `windy', and `snow'.
    These features are one-hot encoded and concatenated to form the context vector.
    \item \textbf{Action set initialization:} For each user, we define a set $\mathcal{A}$ of possible sleep-enhancing activities (\eg, evening walk, meditation, reading).

    \item \textbf{Decision rule:} At each time step $t$, for a given context $x_t$, LinUCB selects the action $a_t$ that maximizes:
    \[
    a_t = \arg\max_{a \in \mathcal{A}} \left( \hat{\theta}_a^\top \mathbf{x}_t + \alpha \sqrt{\mathbf{x}_t^\top \mathbf{A}_a^{-1} \mathbf{x}_t} \right),
    \]
    where:
    \begin{itemize}
        \item The term $\hat{\theta}_a^\top \mathbf{x}_t$ represents the estimated expected reward for action $a$ given context $\mathbf{x}_t$. $\hat{\theta}_a \in \mathbb{R}^d$ is the estimated parameter vector for action $a$, representing the learned relationship between the context and the reward (sleep quality) for that action.
        
        \item The term $\sqrt{\mathbf{x}_t^\top \mathbf{A}_a^{-1} \mathbf{x}_t}$ represents the uncertainty (standard deviation) in the estimate, forming the upper confidence bound. $\mathbf{A}_a \in \mathbb{R}^{d \times d}$ is the covariance matrix (or precision matrix) for action $a$, initialized as a $d \times d$ identity matrix and updated over time. It records the certainty of our estimates.
        
        \item $\alpha$ (>0) is a positive scalar that controls the trade-off between exploration and exploitation. A higher $\alpha$ encourages more exploration.
        
    \end{itemize}
    
    \item \textbf{Action execution:} The system recommends the selected activity $a_t$ to the user.
    
    \item \textbf{Reward observation:} After recommendation, the system observes a reward $r_t$
    (\eg, a normalized sleep score between 0 and 1).
    
    \item \textbf{Update procedure:} We then update the parameters for the suggested action $a_t$ based on the observed reward $r_t$:
    \begin{enumerate}
        \item Update the covariance matrix by incorporating context information at time $t$ to refine the certainty in parameter estimates:
        $\mathbf{A}_{a_t} = \mathbf{A}_{a_t} + \mathbf{x}_t \mathbf{x}_t^\top$.
        
        \item Update the accumulated reward vector for action $a_t$:
        $\mathbf{b}_{a_t} = \mathbf{b}_{a_t} + r_t \mathbf{x}_t$.
        
        \item Recompute the parameter vector for action $a_t$:
        $\hat{\theta}_{a_t} = \mathbf{A}_{a_t}^{-1} \mathbf{b}_{a_t}$.
    \end{enumerate}    
\end{itemize}

\subsection{Sleep Health Chatbot}\label{sec.chatbot}
While quantitative data from users (\eg, sleep, activity, context) is valuable for promoting sleep health, it often overlooks nuanced personal factors such as lifestyle preferences and individual barriers. 
To bridge this limitation, we develop an LLM-powered chatbot that enables natural, flexible interactions. 
By integrating quantitative evidence with behavior change frameworks, the chatbot delivers data-driven and theory-informed advice tailored to users' unique circumstances.
This integrated approach aims to create a positive feedback loop where recommended behavior changes lead to improved sleep outcomes.

\subsubsection{Build Theoretical Framework for Conversation}\label{subsec.chatbot_framework}
We incorporate widely recognized behavior change theories
to inform the chatbot's conversational flow and recommendation strategies.
The use of these theories provides a structured and solid foundation for identifying the determinants of behavior change and selecting appropriate strategies to address them.
They can guide LLMs to generate more relevant and reliable responses.

\textbf{Selection of behavior change theories.}
Following \citet{moullin2020ten}'s recommendations, our framework selection is guided by four key criteria:
1) purpose alignment with behavior change and implementation;
2) individual-level target capability;
3) capability to address implementation elements (\eg, barriers and strategies);
and 4) contextual fit with individual lifestyles and chatbot interventions.
After evaluating several frameworks, 
we choose Theoretical Domains Framework (TDF)~\cite{atkins2017guide} and the Behavior Change Technique (BCT) Taxonomy~\cite{michie2013behavior} because of their comprehensive coverage of behavior determinants and actionable techniques, and their suitability for individual-level interventions.
Specifically,
TDF offers 14 domains for identifying behavior change factors,
such as knowledge, skills, belief about capabilities, and environmental context.
BCT Taxonomy provides 93 specific techniques organized into 14 categories (\eg, goals and planning, feedback and monitoring) for implementing behavior change interventions.
It offers practical methods for translating theoretical understanding into actionable interventions.
This combination allows us to identify behavioral determinants and address them through actionable strategies.

We have also considered alternative frameworks. 
Behavior Change Wheel (BCW)~\cite{michie2011behaviour} combines the COM-B model (Capability, Opportunity, Motivation—Behavior) with intervention functions and policy categories to design comprehensive behavior change interventions.
It emphasizes policy enablers and broader intervention functions that are less directly applicable to our context.
Similarly, we rule out Consolidated Framework for Implementation Research (CFIR)~\cite{damschroder2009fostering} since it highlights contextual factors such as organizational culture and policies.
The Transtheoretical Model (TTM)~\cite{prochaska1997transtheoretical} describes behavior change as a process through six stages, from precomtemplation to termination.
It focuses more on tailoring interventions for transitions of change over extended periods, which is less aligned with our immediate, adaptive interventions.

\textbf{Consolidating theoretical frameworks.}
We first identify TDF domains that are most relevant for sleep behavior.
Some domains like Social Role and Identity and Social Influences are excluded due to the challenges of simulating users' social connections in a chatbot context.
Then, we select BCT categories that align with the identified TDF domains and are practical for a chatbot to deliver.
Finally, we consolidate the concepts derived from TDF and BCT into seven key techniques to inform our chatbot design. 
For example, the ``intention'' and ``goals''
domains from the TDF are merged as one category because they both represent a commitment to achieve a desired outcome. 
The TDF domain ``Behavior Regulation'' is merged with BCT ``Feedback and Monitoring'', both involving objective tracking and management of observed behaviors.
BCT ``Social Support'' category is scoped to TDF ``Emotion'' domain.
The final framework is detailed in \autoref{tab:sleep-techniques}.

\subsubsection{Implementing Data-Driven and Theory-Guided Chatbot}\label{subsec.chatbot_implementation}
We implement a multi-agent chatbot where each agent, powered by LLMs, performs specialized analytical tasks to generate comprehensive, personalized responses. 
They coordinate with each other by custom prompts and rules.

First, \textbf{Agent Coordinator} analyzes incoming messages and conversation history to determine which analytical tasks are needed. It then directs requests to corresponding specialized agents based on classification prompts.
The agents then process these requests:

\begin{itemize}
    \item \textbf{Technique Extraction Agent:} 
    This agent identifies relevant behavior change strategies (see \autoref{tab:sleep-techniques}) for generating responses.
    It uses tailored prompts that fuse users' current messages, conversation history, and behavior change theories to instruct LLM selection of response strategies.

    \item \textbf{Personal Data Agent:} This agent retrieves and analyzes the user's personal data (\autoref{tab:data_collection}), such as sleep patterns and activity levels, that are relevant to the user's messages.
    Given the user's chat messages and daily wearable data (organized as a dataframe) as input, the agent leverages the LLM's capabilities to generate Python code for data retrieval, transform, and analysis.
    Based on the code results and user message, another LLM reports the insights (\eg, average sleep duration over the past week).\footnote{If there are data and processing errors, the system will respond ``I am sorry, I am not able to provide the information at the moment.''}

    \item \textbf{Recommendation Agent} is responsible for delivering context-aware personalized activity suggestions.
    Given the current users' context data (\eg, weather, time of day), our contextual MAB model suggests sleep-enhancing activities (\eg, run, walk). Then, to improve relevance and actionability, the agent prompts LLMs to adjust and tailor the model-generated activities based on the user's current contexts, locations, and situations implied in the messages. For example, it may suggest indoor treadmill running when it is raining outside or walking in the nearby park.
   
\end{itemize}

Finally, the \textbf{Response Agent} synthesizes outputs from all agents into cohesive replies that integrate data insights, personalized recommendations, and behavior change techniques. 
This modular approach decomposes complex chatbot's conversational and analytical reasoning tasks into specialized agents, ensuring responses are relevant, theory-grounded, and adaptive to user's specific circumstances.

\begin{table*}[h]
\centering
\caption{Seven core techniques identified for sleep health improvement.}
\label{tab:sleep-techniques}
\resizebox{.95\textwidth}{!}{
\begin{tabular}{|p{0.2\textwidth}|p{0.4\textwidth}|p{0.4\textwidth}|}
\hline
\textbf{Technique Domain} & \textbf{Definition} & \textbf{Example} \\
\hline
1. Consequences and Reinforcement & Discussing anticipated outcomes of sleep behaviors and providing feedback on user's actions & ``Based on your Oura ring data, your consistent 10 PM bedtime has led to a 15\% increase in your deep sleep. This improvement can enhance your memory and cognitive function.'' \\
\hline
2. Feedback and Monitoring & Tracking sleep patterns and providing users with insights into their progress & ``I notice you've been using your Oura ring consistently. Let's review your sleep efficiency over the past week and identify areas for improvement.'' \\
\hline
3. Goals & Setting clear, achievable sleep-related objectives tailored to the user's current habits and desired outcomes & ``Given your current average of 6 hours of sleep, shall we set a goal to gradually increase this to 7 hours over the next month?'' \\
\hline
4. Knowledge & Providing users with tailored information about sleep health, addressing gaps in their understanding & ``Did you know that exposure to blue light from devices before bedtime can disrupt your melatonin production? Let's discuss some strategies to minimize this effect.'' \\
\hline
5. Environmental Context and Resources & Addressing the user's physical sleep environment and available resources to optimize sleep conditions & ``I see you live in a noisy urban area. Have you considered using a white noise machine to mask disruptive sounds during the night?'' \\
\hline
6. Skills and Capabilities & Teaching users specific techniques to improve their sleep and building their confidence in implementing these strategies & ``Let's practice a simple breathing exercise that can help you relax before bed. Inhale for 4 counts, hold for 7, and exhale for 8. How does that feel?'' \\
\hline
7. Emotional Support & Providing empathy, encouragement, and motivation to support users' sleep improvement efforts & ``I understand that changing sleep habits can be challenging. Remember, every small step you take is progress. How can I help you feel more confident about making these changes?'' \\
\hline
\end{tabular}
}
\end{table*}

\section{System Implementation}\label{sec.sys_implementation}
\name{} is implemented as an interactive web-based application using Streamlit, a Python framework to build data-driven web apps.
Upon user content, the system accesses the user's location data via the browser, which is then used to retrieve local time, weather conditions, and temperature information through the WeatherAPI\footnote{\url{https://www.weatherapi.com/}}. 
This contextual information is used to build the personalized activity recommendation model.
Meanwhile, users' wearable data, including activity, sleep, and physiological data (\eg, heart rate), is accessed through Oura API\footnote{\url{https://cloud.ouraring.com/v2/docs}}.
The chatbot's core functionality is powered by OpenAI's language models. 
GPT-4o-mini is employed for managing general conversation flow, theoretical framework selection, and recommendation model integration and results refinement, ensuring a balance between response quality and speed. 
For the more complex task of retrieving and analyzing wearable data, we leverage GPT-4o to use the Python tool to generate analysis codes and results.
Then, we build the LLM-based multi-agent framework and chatbot by using Python LangChain Library\footnote{\url{https://github.com/langchain-ai/langchain}}. LangChain provides a suite of tools for constructing prompts, managing conversations, and orchestrating the flow between agents and LLMs via prompt chaining. It allows us to modularize the chatbot's functionalities into specialized agents that interact seamlessly.
To enhance user experience, the system implements streamed output, reducing perceived latency and allowing for real-time interaction.
The application is hosted on Amazon Elastic Compute Cloud (EC2), providing scalable computing capacity.

\section{Evaluation}\label{sec.evaluation}

We conducted an in-the-wild deployment study with 16 participants to evaluate our system compared to a baseline chatbot. 
We aimed to answer the following research questions:
\begin{itemize}
    \item How does \name{} impact users' sleep health behaviors and outcomes?
    \item How effective are \name{}'s activity recommendations and theory-guided conversations in motivating behavior change?
    \item What are users' perceived benefits and limitations towards using \name{}?
\end{itemize}

\subsection{Study Design}\label{subsec.study_design}
We employed a within-subjects design to evaluate \name{} against a baseline system. The study lasted for eight weeks and consisted of three phases:
\begin{itemize}
    \item \textbf{Observation (16 days):} Participants maintained their regular sleep and activity patterns, such that we can collect user historical data to initialize our models.
    \item \textbf{\name{} (20 days):} Our chatbot~\jianben{engaged} users in theory-guided and data-driven conversations to motivate users' sleep-enhancing behavior change. 
    \item \textbf{Baseline (20 days):} Participants used a ChatGPT-style chatbot capable of answering general health-related questions, providing general sleep advice, and answering questions about users' health data.
\end{itemize}
To mitigate ordering effects, we counterbalanced the Baseline and \name{} phases across participants. 

\subsection{System Implementation and Setup}
Both \name{} and the Baseline were implemented as web-based applications hosted on AWS EC2, accessible via URL. Key differences between the two systems were:
\begin{itemize}
    \item \name{}: Implemented as described in \autoref{sec.chatbot}, featuring theory-guided conversations and personalized activity recommendations.
    \item Baseline: Followed a similar framework but lacked the theory-guided approach and personalized activity recommendations of \name{}. Specifically, the baseline leverages agent coordinators to decide whether users ask questions about their data. If yes, the data insight agent uses a Python tool to retrieve and analyze the target data subsets and provide a response based on the conversation context.  Otherwise, the chatbot responds to the user message directly.
\end{itemize}
Both systems were initialized with a system prompt to act as a sleep expert, helping users improve sleep quality.

\subsection{Participants}\label{subsec.participants}
We recruited (N=16)
participants (age range: 18-44 years, 5 females, 10 males, 1 prefer not to say) from diverse sources, including email listservs and posters at university, personal connections, and social clubs (\eg, running clubs).
\autoref{tab:demographics} summarizes participant demographics.
They were primarily Asians (9) and Whites (5); most held graduate degrees (14).
Regarding work arrangements,
7 split time evenly between home and office, and 5 worked mainly in the office; almost all resided in urban areas.

Prior to the study, 7 participants did not own or wear a wearable device, and 9 reported using a wearable device for varying periods from less than 6 months to over 2 years.
Participants were highly motivated—with 12 (75\%) very or extremely motivated to improve physical activity and 13 (81.25\%) to improve sleep health.
Sleep patterns varied among participants, with 6 self-identifying as night owls, 5 as average sleepers, 3 as early birds, and 2 as irregular.
The most common bedtimes were between 11 to 12 pm, and the most common wake times were between 6 to 9 am.
This varied sample enabled the evaluation of our sleep health chatbot across different profiles, work arrangements, and sleep habits.

\begin{table}[htbp]
\centering
\caption{Participant demographics.}
\label{tab:demographics}
\small 
\resizebox{\columnwidth}{!}{
\begin{tabular}{p{0.35\columnwidth} p{0.42\columnwidth} r}
\hline
\textbf{Characteristic} & \textbf{Category} & \textbf{N (\%)} \\
\hline
\textbf{Age} & 18 to 24 & 3 (18.75) \\
 & 25 to 34 & 9 (56.25) \\
 & 35 to 44 & 4 (25.00) \\
\hline
\textbf{Gender} & Female & 5 (31.25) \\
 & Male & 10 (62.50) \\
 & Prefer not to say & 1 (6.25) \\
\hline
\textbf{Race/Ethnicity} & Asian & 9 (56.25) \\
 & Black or African American & 1 (6.25) \\
 & Other & 1 (6.25) \\
 & White & 5 (31.25) \\
\hline
\textbf{Occupation} & Full-time employed & 8 (50.00) \\
 & Full-time employed and Student & 2 (12.50) \\
 & Not employed & 1 (6.25) \\
 & Student & 5 (31.25) \\
\hline
\textbf{Wearable Device Usage} & No prior ownership & 6 (37.50) \\
 & Own but never use & 1 (6.25) \\
 & Less than 6 months & 3 (18.75) \\
 & 6 months to 2 years & 3 (18.75) \\
 & Over 2 years & 3 (18.75) \\
\hline
\textbf{Motivation to Improve Physical Activity} & Extremely motivated & 4 (25.00) \\
 & Very motivated & 8 (50.00) \\
 & Somewhat motivated & 4 (25.00) \\
\hline
\textbf{Motivation to Improve Sleep Health} & Extremely motivated & 5 (31.25) \\
 & Very motivated & 8 (50.00) \\
 & Somewhat motivated & 2 (12.50) \\
 & Not very motivated & 1 (6.25) \\
\hline
\textbf{Sleep Pattern} & Night owl & 6 (37.50) \\
 & Average sleeper & 5 (31.25) \\
 & Early bird & 3 (18.75) \\
 & Irregular & 2 (12.50) \\
\hline
\end{tabular}
}
\end{table}

\subsection{Study Procedures}

\subsubsection{Pre-study Preparations}
We took comprehensive measures to ensure ethical conduct and participant readiness.
First, participants were thoroughly informed about the study protocols, such as data collection methods and privacy measures. They signed consent forms detailing the study's purpose, procedures, risks, and benefits.
Participants were instructed on how to interact with the chatbots, asking questions about their sleep and activity data and requesting suggestions. Sample questions were provided to help initiate conversations with both systems.
For data protection, participants were assigned unique identifiers (user ID and password) to access the chatbots.
They were informed about the chatbots' access to their Oura ring metrics and instructed not to input sensitive content. 
Oura rings were provided along with instructions for granting and later revoking API access for data collection.

\subsubsection{Study Phases}
After the observation periods, participants used either the Baseline or \name{} for three weeks. 
Finally, semi-structured interviews (45 minutes) were conducted at the study's end to gather in-depth feedback on both systems.
Throughout the study, participants were asked to wear their Oura ring continuously and carry their iPhones, complete daily questionnaires, participate in post-system questionnaires, and engage in exit interviews at the study's conclusion.

\textbf{Daily questionnaires}
collected users' perceived sleep quality, motivation levels, and responses to chatbot recommendations.
The daily survey included the following key components:
(1) sleep quality rating (5-point scale: ``Terrible'' to ``Excellent''); 
(2) motivation level to improve sleep/activity (5-point scale);
(3) reasons for low motivation (if applicable);
(4) relevance of chatbot recommendations (5-point scale: ``Not Relevant at all'' to ``Very Relevant'');
(5) adherence to recommendations (Yes/No);

\textbf{Post-system surveys}
assessed user experience on a 5-point Likert scale:
(1) technical aspects (e.g., response time, accuracy, user interface design, overall usability);
(2) personalization and relevance of recommendations;
(3) impact on understanding sleep-activity relationships;
(4) motivation and behavior change;
(5) perceived improvements in sleep quality.
Open-ended questions were also collected on behavior changes and system strengths and limitations.

\textbf{Exit interviews}
gathered feedback on
(1) overall impressions of both systems;
(2) perceived impacts on sleep patterns and physical activity;
(3) perceived engagement and usage over time;
(3) most and least helpful features;
(4) challenges in adopting recommendations;
(5) suggestions for improvement;
(6) likelihood of continued use.

\subsection{Data Collection}
We collected the following data daily throughout the study. 

\textbf{Wearable data.}
Sleep and activity data were collected using the Oura ring (see \autoref{tab:data_collection}). The metrics included duration, efficiency, quality, average HRV, and lowest heart rate, stress levels, and readiness scores.
Activity types and scores were also recorded.

\textbf{Questionnaires and interview data.}
We collected daily questionnaires, post-system questionnaires, and exit interview feedback (detailed in \autoref{subsec.study_design}).

\textbf{System usage data.}
We collected data on users' conversation history with the chatbots.
Based on that, we computed the frequency of interactions and length of conversations.

\subsection{Data Security and Ethical Considerations}
All data was securely stored on a university-hosted Amazon Web Services (AWS) server. 
To protect participants’ privacy, personal information was stored separately from research data, with identifying details securely kept on the EZ backup system.
To ensure ethical use of chatbots, 
We implemented guardrails using OpenAI's moderation API\footnote{\url{https://platform.openai.com/docs/guides/moderation}} to detect and block the generation of potentially harmful content.
Regular check-ins were also conducted to address participants' concerns about chatbot responses.

\subsection{Data analysis}
We conducted analyses corresponding to each research question:

\subsubsection{Impact on Sleep Health Behaviors and Outcomes}
\begin{itemize}
\item Paired t-tests to compare sleep and activity metrics (\eg, sleep duration, sleep efficiency, and readiness score) between Baseline and \name{} conditions
\item Descriptive statistics (mean and standard deviation) for stress level distribution comparison
\item Thematic analysis of user responses regarding perceived impacts on sleep patterns and physical activity
\end{itemize}
\subsubsection{Effectiveness of Recommendations and Theory-guided Conversations}
\begin{itemize}
\item Wilcoxon signed-rank test to compare personalization and relevance ratings, and adherence rates between Baseline and \name{} conditions
\item Descriptive statistics for the frequency of behavior change techniques employed by \name{}
\item Thematic analysis of user feedback on the most and least helpful features of \name{}
\item Content analysis of chat logs to identify prevalent behavior change techniques and their implementation
\item Expert feedback on system responses to users' messages
\end{itemize}
\subsubsection{User Engagement and Perceived Benefits and Limitations}
\begin{itemize}
\item Paired t-tests to compare user engagement metrics (ratio of active days, conversation length) between Baseline and \name{} conditions
\item Linear regression analysis to examine engagement trends over time for both systems
\item Wilcoxon signed-rank test to compare user ratings on various aspects of system performance, usability, and perceived benefits between Baseline and \name{} conditions
\item Thematic analysis of interview responses and open-ended survey questions to identify:
\begin{itemize}
\item Perceived benefits and limitations of \name{}
\item Challenges in adopting recommendations
\item Suggestions for improvement
\end{itemize}
\item Descriptive statistics of user satisfaction ratings
\end{itemize}

\section{Results and Analysis}\label{sec.eval_results}

We analyzed quantitative data from user interactions, Oura ring metrics, ratings, and qualitative participant feedback to provide a comprehensive comparison of \name{} and the Baseline.

\begin{table}[ht]
\centering
\caption{Comparative analysis of sleep and activity metrics collected by Oura for Baseline and \name{}.}
\label{tab:wearable_data_comparison}
\small 
\resizebox{\columnwidth}{!}{%
\begin{tabular}{p{0.27\columnwidth} cc cc c c}
\hline
\textbf{Metric} & \multicolumn{2}{c}{\textbf{Baseline}} & \multicolumn{2}{c}{\textbf{\name{}}} & \textbf{P-val} & \textbf{Sig.?} \\
 & \textbf{Mean} & \textbf{SD} & \textbf{Mean} & \textbf{SD} &  &  \\
\hline
Sleep Duration (hrs) & 6.48 & 0.59 & 6.84 & 0.63 & 0.0268 & Yes \\
Sleep Efficiency (\%) & 80.27 & 5.20 & 80.58 & 4.21 & 0.8490 & No \\
Average Breath & 15.49 & 1.17 & 15.50 & 1.10 & 0.5882 & No \\
Average HRV & 53.45 & 19.14 & 49.15 & 19.85 & 0.6200 & No \\
Lowest Heart Rate & 56.99 & 7.64 & 57.21 & 8.32 & 0.9537 & No \\
Sleep Score & 75.28 & 13.23 & 78.34 & 13.77 & 0.4764 & No \\
Activity Score & 71.37 & 14.18 & 75.41 & 13.94 & 0.0268 & Yes \\
Readiness Score & 76.44 & 6.34 & 77.91 & 6.29 & 0.4570 & No \\
\hline
\end{tabular}%
}
\end{table}

\begin{table}[ht]
\centering
\caption{\rev{Post-system questionnaire ratings of Baseline and \name{}. Medians and interquartile ranges (IQR) are reported. Wilcoxon signed-rank test was used for statistical significance testing.}}
\label{tab:comparative_analysis}
\normalsize 
\renewcommand{\arraystretch}{1.2} 
\resizebox{\columnwidth}{!}{%
\begin{tabular}{p{0.38\columnwidth} c c c c c}
\hline
\textbf{Aspect} & \textbf{Baseline} & \textbf{\name{}} & \textbf{W} & \textbf{\textit{p}-val} & \textbf{Sig?} \\
 & \textbf{Med (IQR)} & \textbf{Med (IQR)} &  &  &  \\
\hline
Speed of chatbot & 4.00 (2.00) & 4.00 (1.00) & 5.00 & 0.4922 & No \\
Accuracy of responses & 2.50 (2.25) & 4.00 (0.00) & 20.00 & 0.0375 & Yes \\
User interface design & 4.00 (1.00) & 4.00 (1.25) & 42.00 & 0.7815 & No \\
Ease of accessing data & 4.00 (1.00) & 4.00 (1.00) & 7.50 & 1.0000 & No \\
Overall system usability & 3.50 (1.25) & 4.00 (0.00) & 37.50 & 0.2850 & No \\
Personalized recommendations & 4.00 (2.00) & 4.00 (0.00) & 9.00 & 0.0308 & Yes \\
Feasible recommendations & 3.50 (1.25) & 4.00 (1.00) & 6.00 & 0.1597 & No \\
Understood activity-sleep relation & 3.50 (2.00) & 4.00 (2.00) & 8.00 & 0.0386 & Yes \\
Learned new strategies & 4.00 (1.00) & 4.00 (1.00) & 33.00 & 1.0000 & No \\
Effective use of Oura data & 3.00 (2.00) & 4.00 (1.00) & 4.00 & 0.0126 & Yes \\
Increased motivation & 3.50 (1.25) & 4.00 (1.25) & 8.00 & 0.0126 & Yes \\
Made changes to routine & 4.00 (1.50) & 3.50 (1.00) & 16.00 & 0.4170 & No \\
Noticed sleep improvements & 3.00 (1.25) & 4.00 (1.00) & 11.00 & 0.0187 & Yes \\
Understood daily impact on sleep & 2.50 (2.00) & 3.00 (1.00) & 2.50 & 0.0461 & Yes \\
Increased awareness of patterns & 4.00 (2.50) & 4.00 (2.00) & 20.00 & 0.1177 & No \\
\hline
\end{tabular}}
\end{table}

\subsection{Impact on Sleep Health Activities and Outcomes}

\subsubsection{Sleep Metrics}
\autoref{tab:wearable_data_comparison} shows that participants using \name{} had a significant increase in sleep duration—an average improvement of 22 minutes per night (t = -2.59, p < 0.05).
However, analysis of sleep scores revealed no statistically significant difference (p = 0.47) between \name{} (78.34) and the Baseline (75.28).
Similarly, while sleep efficiency showed a slight increase (80.27\% vs. 80.58\%) and HRV exhibited a minor decrease (53.45 vs. 49.15), these differences were not statistically significant. 
The average breath rate and lowest heart rate during sleep remained relatively constant between the two phases.

\subsubsection{Activity and Readiness}
\name{} significantly increased physical activity scores, improving from 71.37 (SD = 14.18) to 75.41 (SD = 13.94) (t = -2.59, p < 0.05), indicating its ability to motivate participants to engage in more activity. 
Although readiness scores improved slightly (76.44 to 77.91), this change was not significant (t = 0.77, p = 0.457).

\subsection{Effectiveness of Recommendations and Behavior Change Techniques}

\subsubsection{Personalization, Relevance, and Adherence of Recommendations.}
\rev{As shown in \autoref{tab:comparative_analysis},
\name{} significantly outperformed the Baseline in providing personalized recommendations, confirmed by the Wilcoxon signed-rank test (W=9.00, p=0.0308).
Participants rated \name{}’s advice as more relevant to their needs, with significantly higher relevance scores (Median (IQR): 3.81 (0.34) vs. 3.34 (0.35); W = 20.00, p = 0.023).}
Qualitative responses further highlighted \name{}'s improved contextualization, with one user stating, \textit{``I think having the chatbot know the current weather and temperature outside is motivating. It recommended me to go out for a walk in a sunny afternoon, which I sometimes did.''} However, participants felt that the Baseline's suggestions were ``boring'' and ``generic'' which they already knew, such as maintaining 7 hours of sleep, watching coffee intake, etc.

The \name{}'s ability to provide location-context-aware recommendations was particularly appreciated. One user reported, \textit{``The chatbot suggested `a light jog around Four Freedoms Park in Roosevelt Island in the later afternoon, considering the sunny weather and cooling down the temperature.' It felt like the advice was tailored just for me and my surroundings.''}
Some participants noticed that \name{} knew how to adapt their daily routines to the environment.
One user said, \textit{``I remember it was a bit rainy and cold that morning, the chatbot suggested indoor activities like yoga and treadmill rather than outdoor running, which I normally do.''}
Another added, \textit{``Once, it recommended indoor strength exercises with air conditioning due to the high temperature. It felt very considerate.''}

\name{} also demonstrated flexibility in adapting recommendations to users' specific situations. 
A user mentioned that one time the chatbot recommended an outdoor walk, but he was recovering from COVID. He told the chatbot and the chatbot put forward 15-30 minutes of gentle stretching and yoga with some poses to try.
Another one appraised, \textit{``When I mentioned I don't like meditation, the chatbot quickly switched the suggestions to mindful breathing exercises instead. It felt like it really listened to my preferences.''}

This enhanced personalization resulted in higher adherence rates. 
Participants followed through on a median of 47.94\% (IQR: 34.27\%) of \name{}'s recommendations, which is significantly higher compared to a median of 36.71\% (IQR: 15.49\%) for the Baseline (W = 28.00, p = 0.0386).
Users frequently mentioned the relevance and practicality of the recommendations as key factors in their increased adherence.
However, there was no significant difference between the two systems regarding the perceived feasibility of the recommendations.

\begin{figure}[ht]
\centering
\includegraphics[width=\columnwidth]{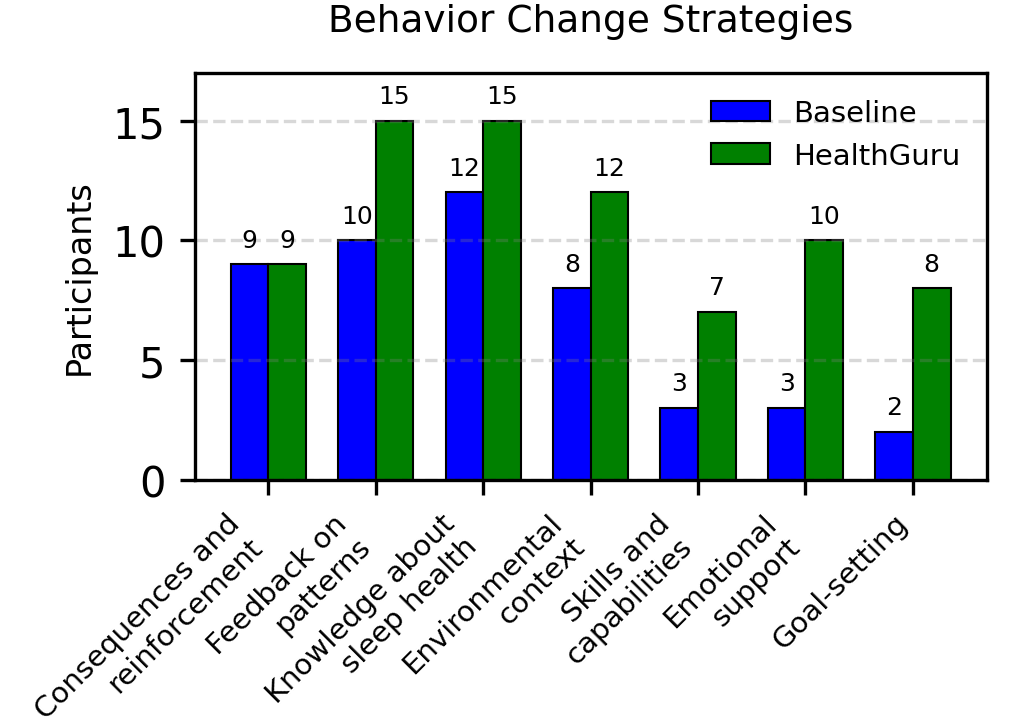}
\caption{Comparison of behavior change strategies covered in Baseline and \name{} systems.}
\label{fig:multiple_choice_comparison}
\end{figure}

\subsubsection{Implementation of Behavior Change Strategies}
Post-system questionnaire analysis results (in \autoref{fig:multiple_choice_comparison}) reveal that \name{} covered a broader range of behavior change strategies compared to the Baseline system. Notably, \name{} showed large improvements in providing feedback on sleep patterns and physical activity (15 vs. 10 participants), addressing environmental context (12 vs. 8), enhancing skills and capabilities for better sleep (7 vs. 3), providing emotional support (10 vs. 3), and facilitating goal-setting (8 vs. 2). 
Both systems performed similarly in addressing consequences and reinforcement (9 participants each).

Analysis of user conversations processed by our multi-agent framework identified seven technique domains, with knowledge being most prevalent (23.8\% of interactions), followed by feedback and monitoring (17.9\%), goals (17.6\%), skills and capabilities (16.9\%), emotion support (9.4\%), consequences and reinforcement (8.2\%), and environment context and resources (6.2\%).
Each strategy was exemplified in \name{}'s responses. For instance, \name{} provided knowledge about potential causes of sleep disturbances and actionable advice:
\begin{quote}
User: \textit{``What are the possible reasons for waking up at midnight?''}\\
\name{}: \textit{``Waking up at midnight can often be linked to stress, sleep disorders, or environmental factors like noise and temperature.''}
\end{quote}

\name{} also encouraged ongoing monitoring and maintaining of positive sleep patterns.
\begin{quote}
\name{}: \textit{``Your sleep efficiency has been consistently high over the past few days, averaging around 91\%. This indicates that you're spending a significant portion of your time in bed actually sleeping, which is a positive sign for your overall sleep quality. Keep monitoring this trend to maintain or improve your sleep health.''}
\end{quote}

In addition, \name{} provided advice to enhance users' skills for improving sleep quality.
\begin{quote}
\name{}: \textit{``Practice a relaxing pre-sleep routine, such as reading or gentle stretching, and incorporate mindfulness or breathing exercises to lower stress levels before bedtime.''}
\end{quote}

Furthermore, qualitative feedback supported the effectiveness of \name{}'s behavior change strategies. Users appreciated the emotional support and personalized feedback. One stated, \textit{``That's great! I'm proud of you for being open to trying.'' }
Another added, \textit{``Feedback on my sleep patterns related to activity... I am trying to be more active. Actually, I just finished a workout. That's been great!''}

\rev{\subsubsection{Expert Feedback}
Three experts (E1-E3) evaluated system responses to users' messages regarding scientific validity, data accuracy, and applicability.
E1 is a health psychologist and the director of a research center for health promotion by physical activity. 
E2 is an assistant professor in sports and exercise science.
E3 is a PhD researcher who has extensive experience (\eg, publications) in sleep and behavior sensing.
We sampled 12 anonymized conversations from study participants using \name{}, along with corresponding wearable data. The samples covered the system's three key capabilities: behavior change techniques, data insights, and activity recommendations, with at least four examples of each\footnote{Individual examples could span multiple categories. They are included in Supplementary Material.}. 

Experts found that \name{} could accurately retrieve and analyze the relevant wearable data according to users' questions.
For example, when asking about ``sleep last night'', \name{} could locate the relevant data attributes, such as sleep duration, efficiency, HRV, and lowest heart rate, on the requested date.
Moreover, E3 added that it was ``user-friendly'' to convert original sleep duration units from seconds to hours.
Experts generally thought the responses were aligned with established sleep and behavior change theories. 
They also appreciated the actionability of \name{}'s activity recommendations, such as ``take a route that passes a nearby park'' while walking to grocery shops and ``walk around the terminal (while at the airport)''.

However, they pointed out areas for improvement.
E3 required a more in-depth scientific interpretation of the users' wearable data (``What an HRV value of 64 means for the user’s sleep'') to facilitate users' understanding beyond simple data summarization.
Some proposed solutions to sleep problems were deemed general and less useful in motivating users to act (``15-30 min yoga''). 
More context information (about ``waking up at midnight'' and ``broken heart'') might be needed to infer users' psychological and physical needs and connect them with appropriate behavior change techniques (E1).
E2 noted that certain conclusions felt less convincing, with ``only one data point used as evidence (in the response).''
}

\begin{table}[ht]
\centering
\caption{Comparison of user engagement metrics between Baseline and \name{} systems.}
\label{tab:user_engagement}
\renewcommand{\arraystretch}{1.2} 
\resizebox{\columnwidth}{!}{%
\begin{tabular}{p{0.31\columnwidth} c c c c}
\hline
\textbf{Metric} & \textbf{Baseline} & \textbf{\name{}} & \textbf{t-stat} & \textbf{\textit{p}-val} \\
 & \textbf{Mean (SD)} & \textbf{Mean (SD)} &  &  \\
\hline
User Engagement & 0.31 (0.19) & 0.39 (0.15) & -2.47 & 0.026* \\
Conversation Length & 4.84 (2.16) & 6.83 (2.67) & -2.18 & 0.045* \\
\hline
\multicolumn{5}{l}{\footnotesize{* Indicates statistical significance at \textit{p} < 0.05.}} \\
\end{tabular}
}
\end{table}

\begin{figure}[ht]
\centering
\includegraphics[width=.95\columnwidth]{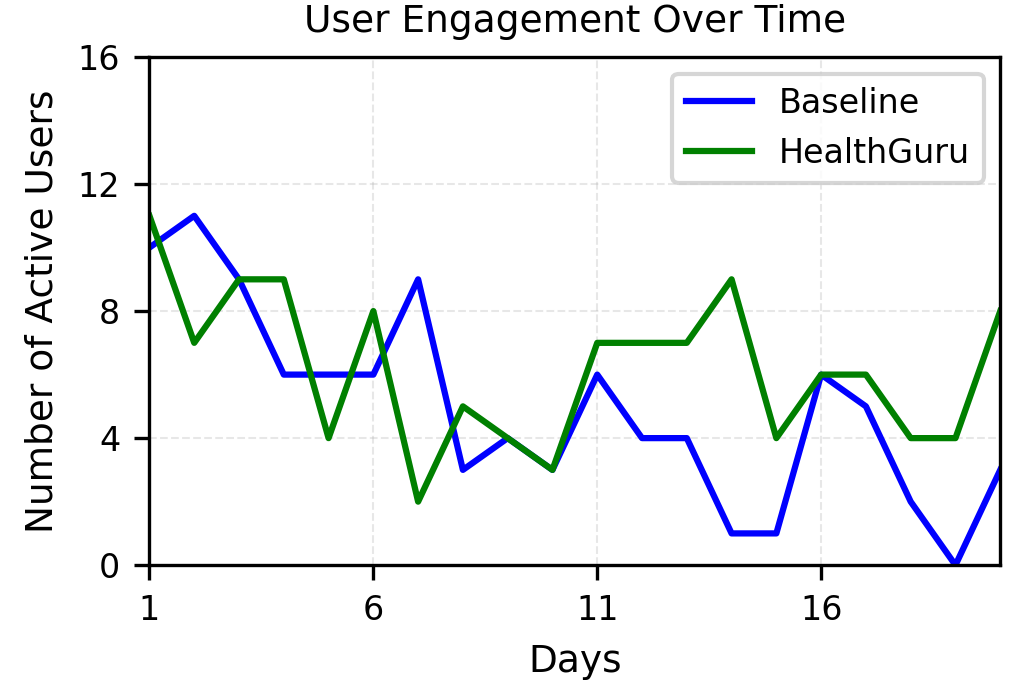}
\vspace{-5mm}
\caption{Comparison of user engagement over time using Baseline and \name{} systems.}
\label{fig:engagement_trends}
\end{figure}

\subsection{User Engagement Analysis}
To evaluate the effectiveness of \name{} compared to the Baseline system, we analyzed key metrics of user engagement (\autoref{tab:user_engagement}): the ratio of active days to total phase days and conversation length. These metrics provided insights into how frequently and deeply users interacted with each system over the course of the study.

\subsubsection{Engagement Metrics}
User engagement, defined as the ratio of active days to total phase days, showed a significant improvement with \name{}. The average engagement ratio increased from 0.31 (SD = 0.19) with the Baseline system to 0.39 (SD = 0.15) with \name{}, a statistically significant improvement (t = -2.47, p = 0.026). 
This result indicates that users interacted with \name{} more consistently over the study period, suggesting enhanced long-term engagement.
The average number of messages in each conversation also increased significantly from 4.84 (SD = 2.16) in the Baseline system to 6.83 (SD = 2.67) with \name{} (t = -2.18, p = 0.045). This suggests that users engaged in longer conversations with \name{}. 
The increased conversation length may reflect users’ higher levels of engagement and \name{}'s ability to provide more comprehensive and tailored responses, which encouraged users to explore health topics in greater detail.

\subsubsection{Engagement Trends Over Time}
To understand how user engagement evolved throughout the study, we analyzed daily active user counts for both systems over the study period (in \autoref{fig:engagement_trends}). 
The Baseline system showed a significant downward trend in engagement with linear regression analysis (slope = -0.018, p = 0.0025, $R^2$ = 0.41). It started with 10 active users, which decreased sharply after the first few days.
In contrast, \name{} demonstrated a more stable engagement pattern. While it started with similar initial engagement (11 active users), it maintained a relatively consistent level of user activity throughout the study period, with periodic fluctuations between 4-9 users. \name{}'s trend analysis showed a much gentler downward slope (-0.0037) that was not statistically significant (p = 0.49, $R^2$ = 0.027), indicating a better sustained user interest over time.

The combination of higher engagement ratios, longer conversations, and more stable daily active users demonstrate \name{}'s effectiveness in maintaining sustained user engagement. This is particularly critical for the long-term success of sleep health interventions, as consistent engagement offers users more opportunities to receive personalized advice, reflect on their sleep patterns, and implement meaningful changes to their routines.

\subsection{System Performance and Usability}
We compared \name{} with the Baseline system in terms of system response, user interface design, and usability metrics (\autoref{tab:comparative_analysis}). 
Our analysis reveal significant improvements in the \name{} system compared to the Baseline system in terms of response accuracy. \rev{The median accuracy ratings of responses improved from 2.50 (IQR = 2.25) to 4.00 (IQR = 1.00), W = 20.00, p = 0.038.} However, there were no significant differences in speed, user interface design, ease of accessing data, or overall system usability between the two systems. This suggests that while \name{} maintained the user-friendly aspects of the Baseline system, it significantly enhanced the quality and relevance of its responses. Qualitative feedback supported this finding. For instance, one user noted, \textit{``The chatbot gave very good advice about my sleep patterns, such as noticing my lack of sleep or lack of exercise. It gave great recommendations personalized to my surroundings and fitness.''}

\subsection{Users' Perceived Benefits and Limitations}
We analyzed user feedback on the perceived benefits and limitations of \name{}, focusing on aspects such as personalization, usefulness, and engagement.

\subsubsection{Perceived Benefits}
\name{} demonstrated significant improvements in several key areas compared to the Baseline system. 

Participants reported a significantly better understanding of sleep-activity relationships when using \name{}. 
\rev{The median score for this aspect increased from 3.50 (IQR: 2.00) to 4.00 (IQR: 2.00) (W = 8.00, p = 0.039). Similarly, users' understanding of how daily activities impact sleep also improved significantly, with median scores rising from 2.50 (IQR: 2.00) to 3.00 (IQR: 1.00) (W = 2.50, p = 0.046).} One user remarked, \textit{``After it told me my sleepiness is due to inactivity rather than bad sleep, I am trying to be more active.''}

\name{} significantly outperformed the Baseline system in the effective use of Oura ring data.
\rev{The median score increased from 3.00 (IQR: 2.00) to 4.00 (IQR: 1.00) (W = 4.00, p = 0.013).} This suggests that \name{} was more effective in leveraging user data to provide meaningful insights.

\name{} also had a notable positive impact on motivation and behavior change. \rev{The median score for increased motivation to improve sleep habits rose from 3.50 (IQR: 1.25) to 4.00 (IQR: 1.25) (W = 8.00, p = 0.013). Additionally, users reported greater sleep improvements after following its advice, with scores increasing from 3.00 (IQR: 1.25) to 4.00 (IQR: 1.00) (W = 11.00, p = 0.019).}

Both systems were effective in increasing users' awareness of their sleep patterns \rev{(median = 4.00 for both)}, but \name{} seemed to inspire more specific behavioral changes. One user reported, 
\textit{``Talking to the chatbot definitely made me more aware of my sleep. Also, there were times I've been paranoid about not sleeping enough, but chatbot told me my sleep score is fine, most likely I am sleepy because of my (low) activity score.''}

\subsubsection{Perceived Limitations and User Suggestions}
Despite the overall positive feedback, users identified several limitations and areas for improvement in \name{}, particularly in data accuracy, actionability of insights, data integration, and user experience.

Some participants reported data accuracy issues in Oura ring. One user noted the sleep score did not align well with their subjective experience. Some mentioned that sometimes the activities were not captured correctly or not detected at all.
One said, \textit{``The app showed that I had a nap, but I was just sitting on chair playing my phone.''}
Moreover, some participants felt the insights could be more actionable and comprehensive. As one user put it, \textit{``The chatbot told me stress is likely a cause, but it didn't give actionable insights. It suggested meditation, but I still had to find my own tutorials.''}
Participants desired system capability to integrate and track factors like daily diet and routines that could influence sleep.

Several usability concerns were raised, including the need to switch between multiple apps, the absence of effective data visualization, and a preference for voice commands over typing. To address these issues, participants suggested developing an integrated platform that seamlessly combines chatbot interactions and data access within a single interface.
In addition, users recommended implementing a dedicated data visualization dashboard to summarize key metrics like stress levels, sleep quality, and their trends over time, making the information easier to interpret. Many participants emphasized the value of voice-based interactions to facilitate system usage, particularly when on the go or in situations where typing is inconvenient.
Participants also highlighted the need of proactive engagement features, such as 
push notifications to highlight progress, track goals, and provide positive reinforcement.

\section{Discussion}\label{sec.discussion}
We introduced \name{}, an LLM-powered chatbot that delivers data-driven, theory-guided sleep health support through a multi-agent framework. This framework features a contextual multi-armed bandit model for adaptive activity recommendations with behavior change strategies in conversational flows.

\subsection{Effectiveness of Data-Driven, Theory-Guided Approach}
An eight-week deployment study with 16 participants demonstrated that \name{} substantially outperformed a baseline system in promoting healthier sleep behaviors. 
By integrating wearable data (\eg, Oura ring metrics) with contextual factors such as time, weather, and location, \name{} delivered recommendations that participants consistently found more personally relevant, context-aware, and actionable.
This improved personalization and contextual relevance led to notably higher adherence rates, a 22-minute increase in average sleep duration, and improved activity scores. 
Participants were more inclined to follow \name{}’s advice—such as taking a light jog in a nearby park—than those provided by the baseline.
Moreover, the inclusion of evidence-based behavior change strategies led to a better understanding of sleep-activity relationships and increased motivation for healthy habits.

\subsection{Balancing Automated Personalization and User Agency}
While participants valued \name{}'s personalized capabilities, the study raises important questions about the balance between automated, AI-driven recommendations and user autonomy. 
Some participants expressed the need to maintain control over their health decisions. 
For example, one mentioned,
\textit{``When it suggests going for a walk because it's sunny, but I feel dehydrated and would prefer to rest.''} This highlights the importance of allowing users to override or adapt system recommendations based on their immediate feelings or circumstances.
Currently, \name{} supports user agency through its conversational interface, enabling users to ask for clarifications or alternatives.
Future approaches can consider interaction designs that allow users to easily specify what aspects of information at what levels of detail should be presented to empower them to make informed decisions.

Moreover, the persuasive nature of personalized recommendations, while effective for behavior change, raises ethical questions about autonomy and informed consent. As one participant noted, \textit{``I find myself following the chatbot's advice more often than not.''} This underscores the need for transparent communication about the system's capabilities and limitations and the importance of framing AI recommendations as suggestions rather than directives.
Future systems should consider implementing more explicit user control features, such as preference settings for recommendation frequency or type, and clear options for users to provide feedback on or override suggestions. Additionally, incorporating periodic reminders of the system's AI nature and encouraging users to consult healthcare professionals for serious concerns could help maintain an appropriate balance between automated support and user agency.

\subsection{Challenges in Sustained Engagement and Behavior Change}
Our study revealed significant challenges in maintaining long-term engagement and translating increased awareness into consistent behavior change. 
\begin{itemize}
    \item Declining novelty: Participants mentioned that initial curiosity and interest in the chatbot's capabilities decreased over time. As one user noted, \textit{``At first, I was curious about what it could suggest, but over time, I felt less motivated to check in with the chatbot.''}
    \item Cognitive load and system interaction barriers: 
    Manual text input introduced a cognitive burden, as evidenced by user feedback requesting voice interactions. Some highlighted the need for better support for handling recurring queries and entry patterns to simplify engagement.
    \item Recommendation fatigue: As participants became familiar with the system's suggestion patterns, they reported decreasing perceived value from recommendations as they began to feel repetitive to users.
    \item Changing life circumstances and competing priorities: Real-life situations, such as travel and recovering from illness, hindered users' ability to engage with the chatbot. 
    \textit{``I knew my sleep was pretty bad lately, but I had to sacrifice my sleep since there is an urgent deadline ahead of me.''}
\end{itemize}

Based on user study results, future systems can consider the following designs.
1) To sustain novelty and interest, systems can introduce adaptive challenge levels that involve users' progress toward their goals, similar to techniques used in gamification. They can also integrate periodic novel content and social features.
2) To improve system interactions, multi-modal interactions (\eg voice commands) and data visualizations should be enabled to enhance natural and convenient data interactions and analytics. Implementation of smart defaults and interaction shortcuts for common queries could simplify engagement.
3) To mitigate recommendation fatigue, systems can continuously expand contextual factors (\eg, seasonal shift) into recommendation logic.
Recommendation timing can be optimized via intelligent notification scheduling based on user receptivity patterns.
4) Systems can support customizable interaction levels, allowing users to adjust the intensity of chatbot engagement based on their current capacity and preferences.

\subsection{Implications for Designing AI-Powered Health Support Systems}
Our study findings suggest three design considerations for future AI-powered health support chatbots.

\par{}\textbf{Emphasizing transparency and reliability to build trust.}
The ability of \name{} to explain the rationale behind its suggestions was highly valued by participants and experts. This can foster trust and motivate adherence to recommendations. By linking advice to personal data and contextual factors, the chatbot helped users understand the benefits of suggested actions. Future health chatbots should prioritize explainability and transparency, clearly communicating how recommendations are generated and how user data is utilized.
This approach aligns with our implementation goals in \name{} and is essential for building long-term user trust.

\par{}\textbf{Applying theory-guided algorithms.}
Operationalizing behavior change theories within data-driven AI systems, as implemented in \name{}, improved response quality and intervention outcomes. The chatbot's use of the TDF and BCT enabled more effective support for behavior change. 
However, careful implementation is required to ensure that theoretical models are appropriately adapted to the AI context. Future systems should continue to incorporate theoretical guidance to enhance data-driven AI methods.

\par{}\textbf{Integration with wearables and health systems.}
Our study reveals opportunities for deeper integration of chatbots with wearables and health systems. 
\begin{itemize}
\item Seamless user experience: Systems should use a unified interface that combines wearable data visualization, chatbot interactions, and health insights. This will address user frustrations with context switching and enhance engagement.
\item Adaptive real-time data analytics: 
Enhancing \name{}'s integration of Oura Ring data, future designs can provide instant, personalized feedback through dynamic dashboards and predictive analytics (\eg, ``Reducing screen time tonight may improve deep sleep by 10\%''). 
These features empower users to make more timely, informed health decisions.
\item Complementing existing health systems: AI chatbots should interpret data from various wearables, providing actionable advice that bridges the gap between users and their health information. This approach can supplement traditional healthcare by supporting daily health management and enabling personalized, continuous interventions.
\end{itemize}

\subsection{Limitations and Future Work}
While \name{} demonstrated promising results in promoting sleep health, our study has several limitations.

\begin{itemize}
    \item \textbf{Data accuracy challenges:} Our system primarily relied on Oura ring, which, while comprehensive and accurate, may still contain 
    missing records, errors, and uncertainty that may impact system performance. Currently, if the system encounters data processing errors, \name{} will respond ``I am sorry, I am not able to provide the information at the moment.'' In addition, users can cross-check raw data tables to judge data quality and system responses. In the future, systems should communicate data quality to users and give more weight to high-confidence data points when generating responses.
    \item \textbf{Contextual factors:} While we incorporated time, location, temperature, and weather into our recommendation agent, other relevant factors such as social context, diseases and medications, or work schedules were not included due to technical and privacy constraints. This limitation may affect the precision of personalized recommendations. 
    \item \textbf{Model reliability evaluation and enhancement:} To improve model reliability, our system adopted multi-agent designs that decompose conversational and analytical tasks into specialized agents to form the final response. The system's reliability was further evaluated through both user and expert judgment. We can scale up the evaluation by
    developing automated fact-checking methods using external knowledge bases~\cite{kotonya-toni-2020-explainable-automated,sarrouti-etal-2021-evidence-based}. Those data resources can also enhance the LLM reliability via retrieval-augmented generation and finetuning.
    \item \textbf{Study sample size and diversity:} Our current user study involved 16 participants (primarily young adults). Expanding the demographic diversity and sample size can improve the generalizability of findings.
    \item \textbf{Study duration:} Longer deployment studies beyond the current eight weeks are needed to assess sustained behavior change and long-term engagement.

\end{itemize}

\section{Conclusions}
We presented \name{}, a novel sleep health chatbot that integrates data-driven insights, behavior change theories, and adaptive recommendations via a multi-agent LLM framework. 
Integrating a contextual multi-arm bandit model, \name{} delivers personalized activity recommendations and theory-grounded conversations to support behavior change. 
Our eight-week deployment study demonstrated significant advantages over a baseline system, with participants showing improved sleep duration, more consistent engagement, and increased motivation to adopt healthy sleep habits.
However, our study also revealed challenges in maintaining long-term engagement and supporting sustained behavior change. 
Future work can improve data visualizations, implement more proactive features, and personalize recommendations to broader individual contexts and preferences.

\begin{acks}
We thank the anonymous reviewers for their feedback. This work was supported by grants from National Science Foundation (2212351, 2212175) and National Institutes of Health (R01AG080991, R01AG07\\6234). Fei Wang is the corresponding author.
\end{acks}

\bibliographystyle{ACM-Reference-Format}
\bibliography{main}



\end{document}